\documentclass{emulateapj}

\usepackage{graphicx}
\usepackage{float}
\usepackage{amsmath}
\usepackage{rotating}
\usepackage{subfigure}
\usepackage{epsfig,floatflt}

\begin{document}

\title{The non-Gaussian Cold Spot in the 3-year WMAP data}

\author{M.\ Cruz\altaffilmark{1}}\affil{IFCA, CSIC-Univ. de Cantabria,  Avda. los 
Castros, s/n,\\ E-39005-Santander,\\Spain}\altaffiltext{1}{Also at Dpto. de F\'{\i}sica 
Moderna, Univ. de Cantabria, Avda. los Castros, s/n, 39005-Santander, Spain}

\email{cruz@ifca.unican.es}

\author{L.\ Cay{\'o}n}
\affil{Department of Physics, Purdue University, 525 Northwestern Avenue,
             West Lafayette, \\ IN 47907-2036, USA}

\email{cayon@physics.purdue.edu}

\author{E.\ Mart\'{\i}nez-Gonz\'alez}
\affil{IFCA, CSIC-Univ. de Cantabria, Avda. los Castros s/n,\\ 39005-Santander,Spain}

\email{martinez@ifca.unican.es}

\author{P.\ Vielva\altaffilmark{2}}\affil{IFCA, CSIC-Univ. de Cantabria, Avda. los 
Castros, s/n,\\ 39005-Santander,Spain}\altaffiltext{2}{Also at Astrophysics Group, 
Cavendish Laboratory, Madingley Road, Cambridge CB3 0HE, UK}

\email{vielva@ifca.unican.es}

\and

\author{J.\ Jin}
\affil{Department of Statistics. Purdue University. 150 N. University Street, West Lafayette,\\ IN 47907-2067}

\email{jinj@stat.purdue.edu}

\begin{abstract}

The non-Gaussian cold spot detected in wavelet space in the WMAP 1--year data, is detected again in the coadded WMAP 3--year data 
at the same position ($b = -57^\circ, l = 209^\circ$) and size in the sky ($\approx 10\degr$). 
The present analysis is based on several statistical methods: kurtosis, maximum absolute temperature, 
number of pixels below a given threshold, volume and Higher Criticism. 
All these methods detect deviations from Gaussianity in the 3--year data set 
at a slightly higher confidence level than in the WMAP 1--year data.  
These small differences are mainly due to the new foreground reduction technique and not to 
the reduction of the noise level, which is negligible at the scale of the spot. 
In order to avoid \emph{a posteriori} analyses, we recalculate for the WMAP 3--year data the
significance of the deviation in the kurtosis.
The skewness and kurtosis tests were the first tests performed with wavelets for the WMAP data. 
We obtain that the probability of finding an at least as high deviation in Gaussian simulations is $1.85\%$.
The frequency dependence of the spot is shown to be extremely flat.
Galactic foreground 
emissions are not likely to be responsible for the detected deviation from Gaussianity.

\end{abstract}


\section{Introduction}

The Cosmic Microwave Background (CMB) is at the moment the most useful tool in 
the study of the origin of the universe. A precise knowledge of its power 
spectrum constrains significantly the values of the cosmological parameters which 
determine the cosmological model. 
The 1--year Wilkinson Microwave Anisotropy Probe data (WMAP, Bennett et al. 2003a),
measured the anisotropies of the CMB
with unprecedented accuracy, finding that the standard model fits these data.
A flat $\Lambda$--dominated Cold Dark Matter ($\Lambda$CDM) universe with standard 
inflation explains
most of the observations confirming the widely accepted concordance model.
According to standard inflation, the temperature anisotropies of the CMB are predicted
to represent a homogeneous and isotropic Gaussian random field on the sky.
A first Gaussianity analysis found the data to be compatible with Gaussianity 
(Komatsu et al. 2003).

Several  non--Gaussian signatures or asymmetries were detected 
in the 1--year WMAP data in subsequent works. A variety of methods were 
used and applied in real, harmonic and wavelet space: low multipole alignment statistics 
(de Oliveira--Costa et al. 2004, Copi et al. 2004, 2005, Schwarz et al. 2004, 
Land \& Magueijo 2005a,b,c, Bielewicz et al. 2005, Slosar \& Seljak 2004); 
phase correlations (Chiang et al. 2003, Coles et al. 2004); hot and cold spot 
analysis (Larson \& Wandelt 2004, 2005); local curvature methods (Hansen et al. 2004, 
Cabella et al. 2005); correlation functions (Eriksen et al. 2004a, 2005, Tojeiro et al. 2005); 
structure alignment statistics (Wiaux et al. 2006); multivariate analysis 
(Dineen \& Coles 2005); Minkowski functionals (Park 2004, Eriksen et al. 2004b); 
gradient and dispersion analyses (Chyzy et al. 2005); and several statistics 
applied in wavelet space (Vielva et al. 2004, 
Mukherjee \& Wang 2004, Cruz et al. 2005, 2006, McEwen et al. 2005a and 
Cay\'on, Jin \& Treaster 2005).

The recently released 3--year WMAP data with higher signal to noise ratio 
is key to confirm or disprove all these results. 

In the 3--year papers, the WMAP team (Hinshaw et al. 2006) re-evaluates potential sources of 
systematic errors and concludes that the 3--year maps are consistent with the 1--year
maps. The exhaustive polarization analysis enhances the confidence on the accuracy of the 
temperature maps.
The $\Lambda$CDM model continues to provide the best fit to the data.

Spergel et al. (2006) perform a Gaussianity analysis of the 3--year data. No 
departure from Gaussianity is detected based on the
one point distribution function, Minkowski functionals, the bispectrum and the trispectrum of 
the maps. The authors do not re-evaluate the other statistics showing asymmetries or non--Gaussian 
signatures in the 1--year data.

The aim of this paper is to check the results of Vielva et al. (2004), Cruz et al. (2005), Cay\'on, 
Jin \& Treaster (2005) and Cruz et al. (2006), (hereafter V04, C05, CJT and C06 respectively) 
with the recently released WMAP data.
All these analyses were based on wavelet space. In particular the data 
were convolved with the Spherical Mexican Hat Wavelet (SMHW). 
Convolution of 
a CMB map with 
the SMHW at a particular wavelet scale increases the signal to noise ratio at that scale. Moreover, 
the spatial location of the different features of a map is preserved.

V04 detected an excess of kurtosis in the 1--year WMAP data compared to 10000 Gaussian 
simulations. This excess occurred at wavelet scales around $5\degr$ (angular size in the sky of 
$\approx 10\degr$). The excess was found to be localized in the 
southern Galactic hemisphere.
A very cold spot, called \emph{the Spot}, at galactic coordinates 
($b = -57^\circ, l = 209^\circ$),
was pointed out as the possible source of this deviation. 

C05 showed that indeed \emph{the Spot}
was responsible for the detection. The number of cold pixels below several thresholds
(cold Area) of \emph{the Spot} was unusually high compared to the spots appearing in the simulations.
Compatibility with Gaussianity was found when masking this spot in the data. The minimum temperature of 
\emph{the Spot} was as well highly significant.

C06 confirmed the robustness of the detection and analysed the morphology and the foreground 
contribution to \emph{the Spot}. \emph{The Spot} appeared statistically robust in all the
performed tests, being the probability of finding a similar or bigger spot in the Gaussian 
simulations less than 1\%. The shape of \emph{the Spot} was shown to be roughly circular, using
Elliptical Mexican Hat Wavelets on the sphere. Moreover the foreground contribution in the region
of \emph{the Spot} was found to be very low. \emph{The Spot} remained highly significant independently 
of the used foreground reduction technique. In addition the frequency dependence of 
\emph{the Spot} was shown to be extremely flat.
Even considering large errors in the foreground 
estimation it was not possible to explain the non-Gaussian properties of \emph{the Spot}.

CJT applied Higher Criticism statistics (hereafter HC) 
to the 1--year maps 
after convolving them with the SMHW. This method provided a direct 
detection of
\emph{the Spot}. The HC values appeared to be higher than 99\% of the Gaussian 
simulations.

Note that although \emph{the Spot} has not been detected in real space, this structure exists
but is hidden by structures at different scales. The convolution with the SMHW at the appropriate scale, 
amplifies \emph{the Spot}, making it more prominent.

Several attempts have been made in order to explain the non-Gaussian nature of this cold spot.
Tomita (2005) suggested that local second--order gravitational effects could produce 
\emph{the Spot}. Inoue \& Silk (2006) considered the possibility of explaining \emph{the Spot}
and other large scale anomalies by local compensated voids.
Jaffe et al. (2005a) and Cay\'on et al. (2006) assumed an anisotropic Bianchi VII$_h$ model showing that it could explain
the excess of kurtosis and the HC detection as well as several large scale anomalies. 
On the other hand, 
McEwen et al. (2005b) still detect non-Gaussianity in the Bianchi corrected maps.
Jaffe et al. (2005b) proved the 
incompatibility of the extended Bianchi models including the dark energy term with the 1--year data.
Adler et al. (2006), developed a finite cosmology model which would explain \emph{the Spot} and
the low multipoles in the angular power spectrum. 
Up to date there are no further 
evidences of the validity of any of the above suggested explanations.

Our paper is organized as follows. We discuss the changes in the new WMAP
data release and the simulations in section $\S$2. The analysis using all the mentioned estimators is described 
in section $\S$3.  In section $\S$4, the significance of our findings is discussed. 
We analyse the frequency dependence of \emph{the Spot} in section $\S$5, and 
our discussion and conclusions are presented in sections $\S$6 and $\S$7.

\section{WMAP 3--year data and simulations}

The WMAP data are provided at five frequency-bands, namely K--band (22.8 GHz, one  receiver), 
Ka--band (33.0 GHz, one  receiver), Q--band (40.7 GHz, two  receivers), V--band 
(60.8 GHz, two  receivers) and W--band (93.5 GHz, four  receivers). 
Foreground cleaned maps for the Q, V, and W channels are also available at the 
Legacy Archive for Microwave BAckground Data Analysis (LAMBDA) web site \footnote{http://lambda.gsfc.nasa.gov}.

Most of the 1--year Gaussianity analyses were performed using the WMAP combined, foreground 
cleaned Q--V--W map 
(hereafter WCM; see Bennett et al. 2003a). CMB is the dominant signal at these bands and 
noise properties are well defined for this map. 
The de-biased Internal Linear Combination map, (DILC) proposed
by the WMAP team, estimates the CMB on the whole sky. However its noise properties are complicated
and regions close to the Galactic plane will be highly contaminated by foregrounds.
Chiang, Naselsky \& Coles (2006) find evidences for the foreground contamination of the DILC.
Therefore we will still use the more reliable WCM in the 3--year data analysis.
%
\begin{figure*}
  \begin{center}
    \includegraphics[width=16cm]{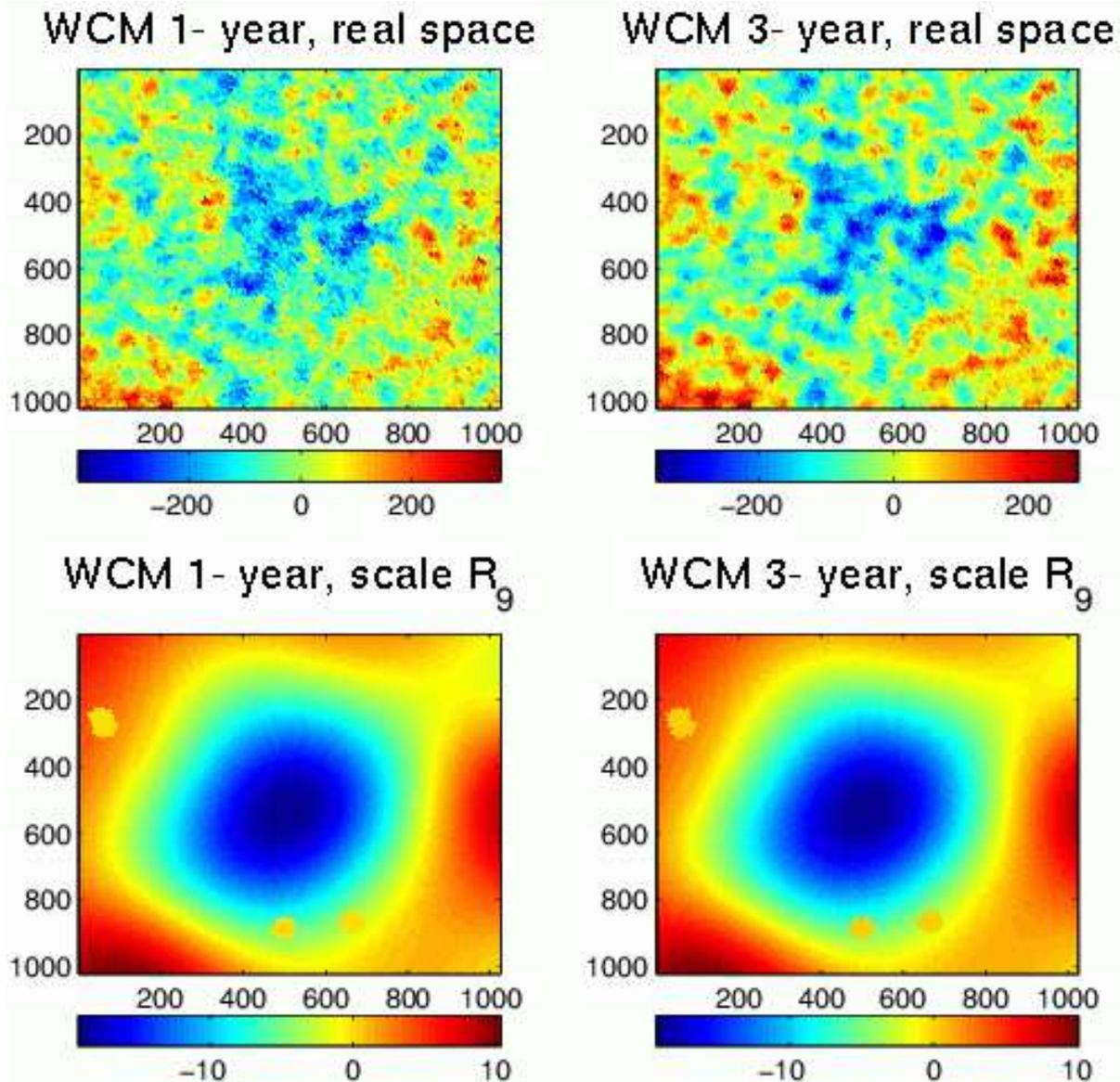}
  \end{center}
  \caption{Image showing an azimuthal projection of a $22\degr \times 22\degr$ patch from the WCM 
    HEALPix map with resolution nside = 256, centered on \emph{the Spot} and in $\mu$K. In the first row 
    we have the 1--year and 3--year images of \emph{the Spot} in real space, whereas in the second
    row \emph{the Spot} is shown at wavelet scale $R_9$. The image is divided in $1024 \times 1024$
    pixels and the y-axis is oriented in the Galactic north-south direction.}
\label{fig:image}
\end{figure*}
\begin{figure}
  \begin{center}
    \includegraphics[width=84mm,height=84mm]{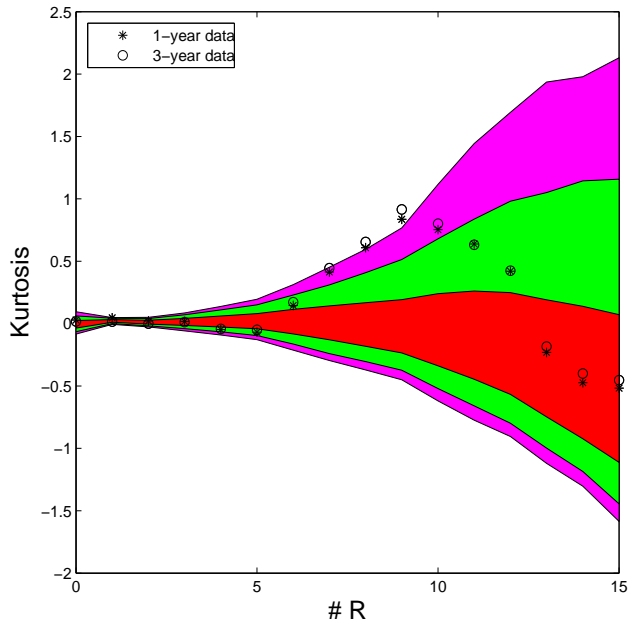}
  \end{center}
\caption{WCM kurtosis values for the 1--year (asterisks) and the 3--year data (circles). 
    The acceptance intervals for the 32\% (inner), 5\% (middle) and 1\% (outer) significance levels,
    given by the 10000 simulations are also plotted.}
\label{fig:kurtosis}
\end{figure}
\begin{figure*}
  \begin{center}
    \includegraphics[width=16cm,height=84mm]{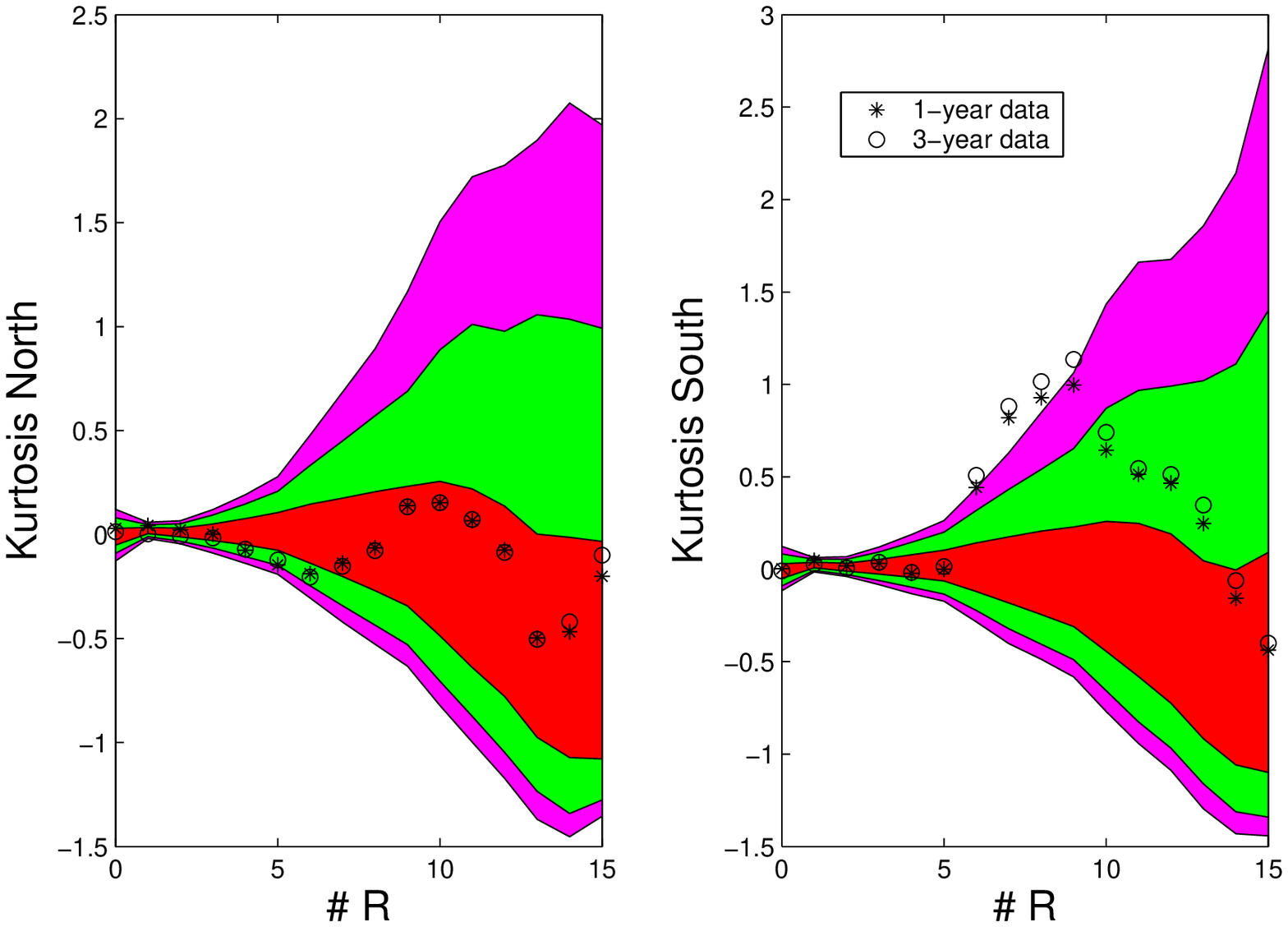}
  \end{center}
  \caption{As in Figure~\ref{fig:kurtosis} but for the northern (left plot) and southern (right plot)
    Galactic hemispheres.}
  \label{fig:kurt_NS}
\end{figure*}

Hinshaw et al. (2006) describe some changes in the 3--year temperature analysis with respect to the
1--year one.
Coadding the three years of observations reduces the instrumental noise. The 3--year maps have 
$\approx 3$ times lower variance.
Refinements in gain calibration and beam response models have been implemented
and a new foreground reduction technique has been used. The latter seems to provide a better correction than the one applied to the first year data.
As discussed in C06 the Galactic foreground estimation is a very important issue in Gaussianity
analyses. 
The exclusion masks defined by Bennett et al. (2003b) have not been modified, except for 
the inclusion of 81 new point sources in the kp0 mask. This mask excludes the highly contaminated
pixels close to the Galactic plane. 

Despite these changes the 3--year maps have been found to be consistent with the 1--year maps by the
WMAP team.

V04 and C05 performed a very careful analysis in order to study the power spectrum and noise
dependence of the kurtosis and cold Area estimators. Considering different power spectra
within the 1$\sigma$ error band of the 1--year data, the differences in the significance of 
the kurtosis were found to be negligible (see Figure 11 in V04). The Area of a particular spot was neither affected by the power spectrum 
(see section 5.3 in C05). The results were almost noise independent.
The convolution with the SMHW 
reduces considerably the noise contribution. 
Even if similar results are expected,  we perform 10000 Gaussian simulations of the 3--year coadded data following
the same steps as for the 1--year simulations. The only differences between the 3--year and the 1--year simulations
are a lower noise contribution and a very slight variation in the power spectrum used to
generate the simulations.
For a detailed description of the simulation pipeline, see section 2 of V04.

We will use all these maps in the HEALPix pixelisation scheme (G\'{o}rski et al. 2005) \footnote{http://www.eso.org/science/healpix/}
with resolution parameter Nside = 256.

\section{Analysis}

Our aim in this section is to repeat the same tests performed in V04, C05, CJT and C06
but with the 3--year data. Then we will compare the new results to the old ones. One can see the region of \emph{the Spot} in real 
and wavelet space at scale $5\degr$ for both releases of the WMAP data in Figure~\ref{fig:image}. In real space the 3--year data image
appears clearly less noisy, whereas the wavelet space images present only very small differences.

In V04, data and simulations were convolved with the SMHW at 15 scales, 
namely ($R_1 = 13.7$, $R_2 = 25$, 
$R_3 = 50$, $R_4 = 75$, $R_5 = 100$, $R_6 = 150$, $R_7 = 200$, $R_8 = 250$, $R_9 = 300$, 
$R_{10} = 400$, $R_{11} = 500$, $R_{12} = 600$, $R_{13} = 750$, $R_{14} = 900$ and 
$R_{15} = 1050$ arcmin). 
The SMHW optimally enhances some non-Gaussian signatures on the sphere
(Mart{\'\i}nez--Gonz{\'a}lez et al. 2002) and has the following expression:
\begin{equation}
\label{eqSMHW}
   \Psi_S(y,R) = 
\frac{1}{\sqrt{2\pi}N(R)}{\Big[1+{\big(\frac{y}{2}\big)}^2\Big]}^2
  \Big[2 - {\big(\frac{y}{R}\big)}^2\Big]e^{-{y}^2/2R^2},
\nonumber
\end{equation}
where $N(R)$ is a normalisation constant: 
$N(R)\equiv R\sqrt{1 + R^2/2 + R^4/4}$.
The distance $y$ on the tangent plane is related to the polar 
angle ($\theta$) as: $y\equiv 2\tan{\theta/2}$.

We will use the same 15 scales in our present analysis, considering
those estimators where non-Gaussianity was found in the 1--year data, namely kurtosis,
Area, $Max$, HC and a new one, the volume.
The definitions of each estimator will be given in the following subsections.
Analyses were also performed in real space, which will be referred as wavelet scale zero.
In real space, the data are found to be compatible with Gaussian predictions

In the following subsections we will give the upper tail probabilities of the data at one particular scale.
The upper tail probability is the probability that the relevant statistic
takes a value at least as large as the
one observed, when the null hypothesis is true.

In section $\S$4 we will give a more rigorous measure of the significance, 
considering the total number of performed tests to calculate the $p$-value of \emph{the Spot}. 
The $p$-value is the probability that the relevant statistic
takes a value at least as extreme as the
one observed, when the null hypothesis is true. 
In our case, the null hypothesis is the Gaussianity of the temperature fluctuations.

\subsection{Kurtosis}

Given a random variable $X$, the kurtosis $\kappa$ is defined as 
$\kappa(X) = \frac{E[X^4]}{(E[X^2])^2} -3$.
In V04 the kurtosis of the wavelet coefficients was compared
to the acceptance intervals given by the simulations. In Figure~\ref{fig:kurtosis} the kurtosis of 
the 1--year data are represented by asterisks and the 3--year data by circles. 
Hereafter we will use these symbols to represent 1--year and 3--year data.
Both are plotted versus the 15 wavelet scales. Scale 0 corresponds to real space. 
The acceptance intervals given by the simulations will be
plotted in the same way in all figures: the 32\% interval corresponds to the inner band, the 5\% interval to the 
middle band and the 1\% acceptance interval, to the outer one. 
As expected, the acceptance intervals remain almost unchanged with respect to those obtained from 1--year 
simulations. This will happen as well for all the other estimators.
The 3--year kurtosis values follow the same pattern as the 1--year ones, confirming the initial
results.
However there are slight differences at the scales where the deviation is detected, being
the kurtosis even higher in the 3--year data. The most significant deviation from the Gaussian
values, occurs at scale $R_9 = 5\degr$. 
In Table~\ref{table:kurtosis} we list the kurtosis values at scale $R_9$,
considering the 1--year data as published in 2003, the 1--year data release applying the changes
in the data analysis described in Hinshaw et al.(2006), and the coadded 3--year data.
The biggest difference is found between both releases of the 1--year data. The kurtosis
value of the 1--year data increases $\approx 7\%$. This may be due to the new foreground 
reduction technique. As expected the noise reduction due to coadding
the three years of observations, implies a much lower increase in the kurtosis, since
the noise contribution in wavelet space is very small.
The upper tail probabilities (i.e. the probabilities of obtaining  higher or equal values
assuming the Gaussian hypothesis) are given in the right column of Table~\ref{table:kurtosis}.
Hereafter we will compare the first release of the 1--year data with the 3--year data.
\begin{deluxetable}{ccc}
\tablewidth{0pt}
\tablecaption{Kurtosis values at scale $R_{9}$\label{table:kurtosis}} 
\tablecomments{Kurtosis values of different WCM versions at scale $R_{9}$. The right column gives
	the probability of obtaining a higher or equal value in Gaussian simulations.}
\tablecolumns{3}
\tablehead{  Data    & kurtosis  & probability\tablenotemark{a} }
\startdata
1--year data (2003)  & 0.836   &   0.38\%    \\
1--year data (2006)  & 0.895   &   0.28\%    \\
3--year data         & 0.915   &   0.23\%    \\
\enddata
\end{deluxetable}
%
%

Analysing both Galactic hemispheres separately, we obtain the results presented in 
Figure~\ref{fig:kurt_NS}.
Again the kurtosis follows the same pattern as in the 1--year results. As expected, the deviation 
appears only in the southern hemisphere and it is slightly higher in the 3--year data. 
The upper tail probability obtained in
V04 was 0.11\% at scale $R_7$ in the southern hemisphere, whereas now we have 0.08\% again at 
scale $R_7$. The deviation from Gaussianity is localised in the southern hemisphere because
\emph{the Spot} is responsible for it (see C05).

\subsection{Maximum statistic}

\begin{figure}
  \begin{center}
    \includegraphics[width=84mm,height=84mm]{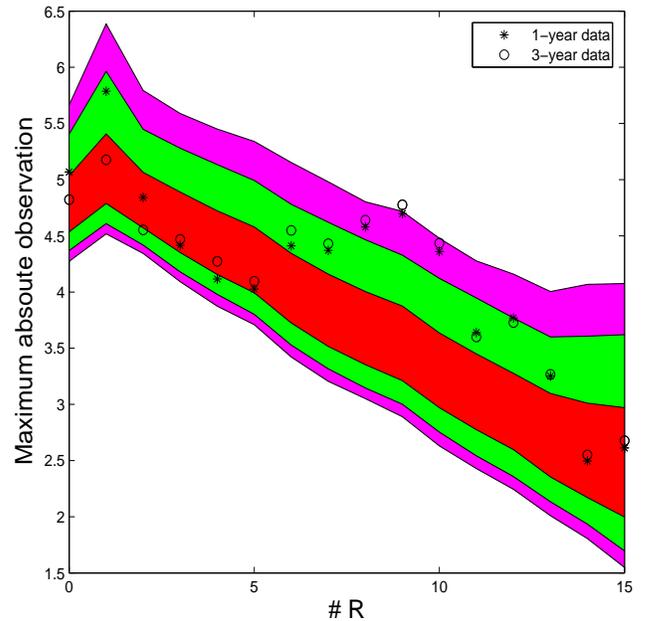}
  \end{center}
  \caption{Maximum absolute observation versus the 15 wavelet scales. Again the circles represent
  the 3--year data and the asterisks the 1--year data. The bands represent the acceptance intervals
  as in previous figures.}
  \label{fig:Max}
\end{figure}

Given $n$ individual observations $X_i$, $Max$ is defined as the largest (absolute) observation :
$$
Max_n   =  max\{ |X_1|, |X_2|, \ldots, |X_n|\}.
$$
The very cold minimum temperature of \emph{the Spot}, was shown to deviate from the Gaussian behaviour in
V04. In this work and in C05, C06 the minimum temperature estimator was used to characterise \emph{the Spot}
whereas in CJT the chosen estimator was $Max$. As $Max$ is a classical and more conservative estimator,
we will use it in the present paper instead of the minimum temperature. 
Our $n$ observations correspond to values in real or wavelet space (normalized to zero mean and dispersion
one). 
\emph{The Spot} appears to be the
maximum absolute observation of the data at scales between 200 and 400 arcmin. In Figure~\ref{fig:Max},
the 1--year and 3--year WMAP data values of $Max$ are compared to those obtained from the simulations.
As for the kurtosis, both data releases show very similar results. The data lie outside the 1\% 
acceptance interval at scales $R_9$ and $R_{10}$. The 3--year data show slightly higher values than
the 1--year data at these scales. In particular, the upper tail probability for the 1--year data was
0.56\%, whereas for the 3--year data we obtain 0.38\% at scale $R_9$.
\begin{figure*}
\begin{center}
\includegraphics[width=16cm,height=84mm]{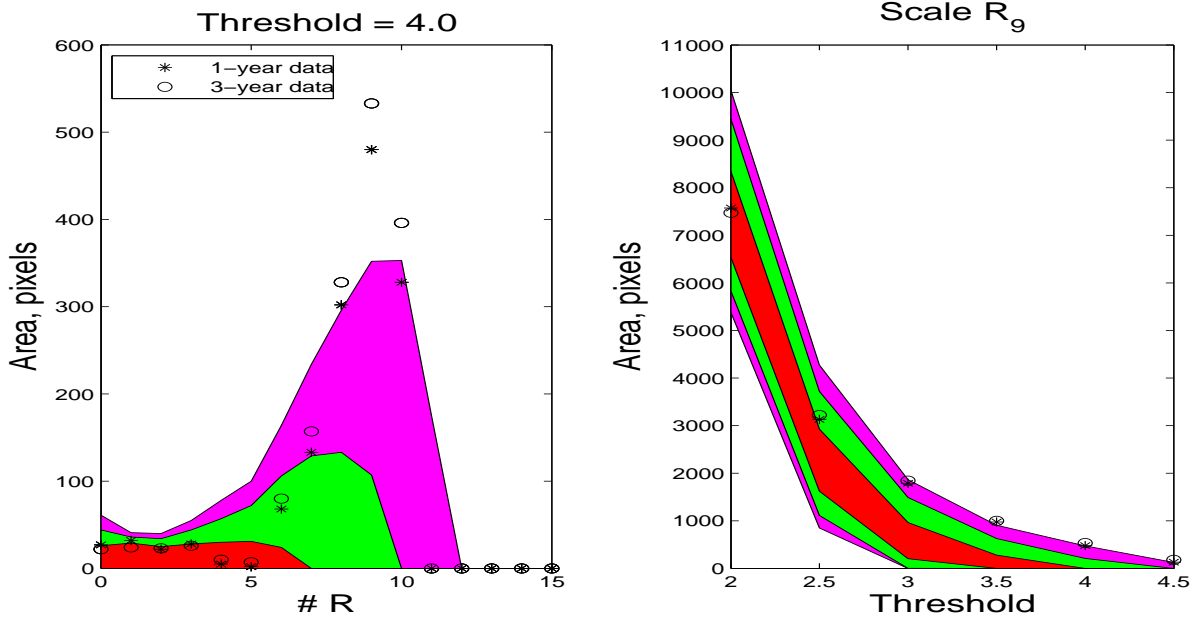}
\end{center}
\caption{The left panel shows the cold Area in pixels, at threshold 4.0 versus the number 
  of the scale. In the right panel the cold Area is represented versus the thresholds, while 
  the scale is fixed at $R_9$. As in previous figures the asterisks represent the 1--year
  and the circles the 3--year data. The bands represent the acceptance intervals as 
  in Figure~\ref{fig:kurtosis}}
\label{fig:Cold_Area}
\end{figure*}
\begin{figure}
\begin{center}
\includegraphics[width=84mm,height=84mm]{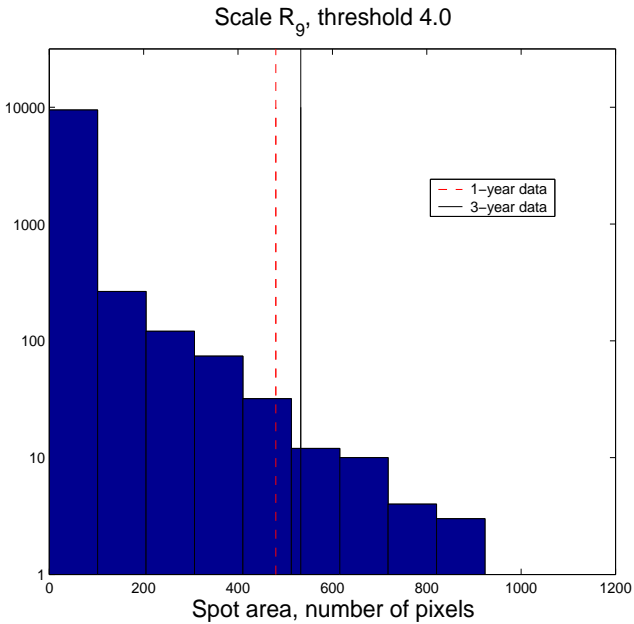}
\end{center}
\caption{Histogram of all biggest spots of the simulations at threshold 4.0 and scale $R_{9}$. 
The dashed vertical line represents \emph{the Spot} in the 1--year data and the solid one represents 
\emph{the Spot} in the 3--year data.}
\label{fig:Histbig}
\end{figure}
\begin{deluxetable}{ccc}
\tablewidth{0pt}
\tablecaption{Upper tail probabilities for the Area of \emph{the Spot} at scale $R_{9}$\label{table:Big spots}}
\tablecolumns{3}
\tablehead{ threshold & probability 1--year data & probability 3--year data }
\startdata
	  3.0 & 0.68\% & 0.63\% \\ 
	  3.5 & 0.36\% & 0.37\% \\ 
	  4.0 & 0.34\% & 0.27\% \\ 
	  4.5 & 0.44\% & 0.35\% \\ 
\enddata
\end{deluxetable}
\begin{deluxetable}{ccc}
\tablewidth{0pt}
\tablecaption{Upper tail probabilities for the volume of \emph{the Spot}, scale $R_{9}$\label{table:Volume}}
\tablecolumns{3}
\tablehead{ threshold & probability 1--year data & probability 3--year data }
\startdata
	  3.0 & 0.51\% & 0.45\% \\ 
	  3.5 & 0.33\% & 0.38\% \\ 
	  4.0 & 0.32\% & 0.27\% \\ 
	  4.5 & 0.44\% & 0.35\% \\ 
\enddata
\end{deluxetable}

\subsection{Area}

We define the hot Area as the number of pixels above a given threshold $\nu$
and the cold Area as the number of pixels below a given threshold $-\nu$. 
The threshold is given in units of the dispersion of the considered map. 

In C05 the total cold Area of the 1--year data was found to deviate from the Gaussian behaviour
at scales $R_8$ and $R_9$ and thresholds above 3.0 (see Figures 1 and 2 in C05).

C05 found that the large cold Area of \emph{the Spot} was responsible for this deviation.
Such a big spot was very unlikely to be found under the Gaussian model at several thresholds 
(see Table 2 of C05). 

In the present paper we will define the Area as the maximum between hot and cold Area at a given threshold
and scale. As for the $Max$ estimator, we obtain in this way a more conservative estimator since
\emph{the Spot} will be compared to the biggest spot in each simulation no matter if it is a cold
or a hot spot. 

However the Area still deviates from the Gaussian behaviour as can be seen in Figure~\ref{fig:Cold_Area}.
The most significant deviation is again found at scale $R_9$ and thresholds above 3.0.

Figure~\ref{fig:Histbig} shows the histogram of the biggest spot of each simulation compared to
the 1--year and 3--year Area of \emph{the Spot} at scale $R_{9}$ and threshold 4.0.
\emph{The Spot} is more prominent in the 3--year data and only very few simulations show bigger spots.
The upper tail probabilities obtained at scale $R_{9}$ for
1--year and 3--year data are presented in table ~\ref{table:Big spots}. 
As in the previous estimators, the 3--year
data are in general slightly more significant.
The new and more conservative definition of the Area estimator reduces the upper tail probability of \emph{the Spot}
although it is still widely below 1\%.

\subsection{Volume}

From the previous subsections we know that \emph{the Spot} is extremely cold and it has a large Area at thresholds
above 3.0. The best estimator to characterise \emph{the Spot} would be therefore the 
volume. Hence we define
the volume referred to a particular threshold as the sum of the temperatures of the pixels conforming 
a spot at this threshold.
In Table~\ref{table:Volume} we compare the probability of finding a spot with higher or equal Volume as
the data, assuming the Gaussian hypothesis. The values are very similar to those obtained for the 
Area estimator. Values for the Volume are slightly more significant and they show less variations with the threshold.
\begin{figure}
\begin{center}
\includegraphics[width=84mm,height=84mm]{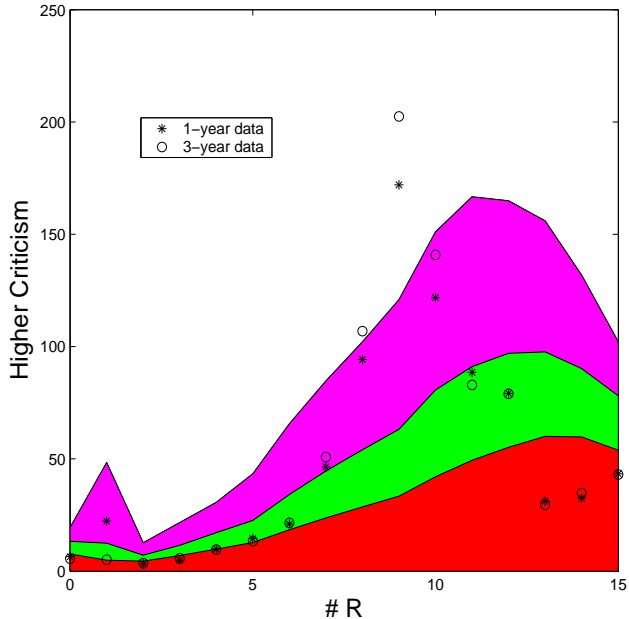}
\end{center}
 \caption{Higher Criticism values of the 1--year WCM (asterisks) and the 3--year WCM (circles). 
   The acceptance intervals are plotted as in previous figures.}
\label{fig:HC_1}
\end{figure}
\begin{figure}
\begin{center}
\includegraphics[width=84mm]{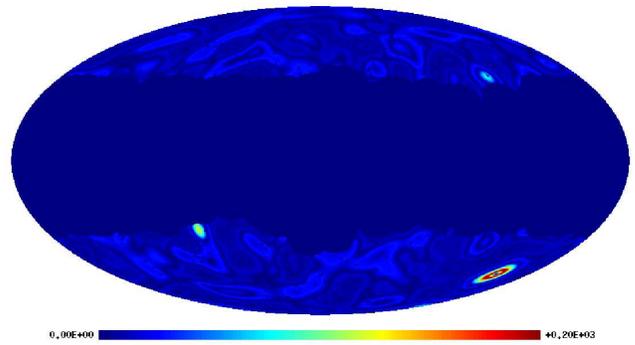}
\end{center}
\caption{Higher Criticism of the 3--year WCM at scale $R_9$.}
\label{fig:HC_2}
\end{figure}
\begin{figure}
\begin{center}
\includegraphics[width=64mm]{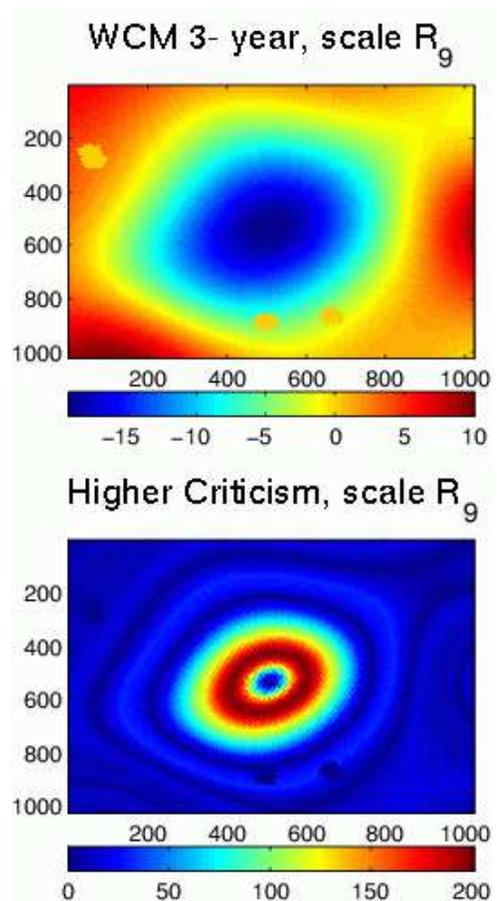}
\end{center}
\caption{Image projected as in Figure~\ref{fig:image}, showing the 3--year WCM map 
(upper panel) and the Higher Criticism map (lower panel), both at scale $R_9$.}
\label{fig:HC_3}
\end{figure}

\subsection{Higher Criticism}

The HC statistic proposed by Donoho \& Jin (2004) was designed to 
detect deviations from Gaussianity that are caused by either a few extreme 
observations or a small proportion of moderately extreme observations. 
Moreover, the statistic provides a direct method to locate these extreme observations by means 
of HC values calculated at every individual data point.

For a set of $n$ individual observations $X_i$
from a certain distribution ($X_i$ normalized to zero mean and dispersion
one), $HC$ is defined as follows. The $X_i$ observed
values are first converted into $p$-values:
$p_{(i)} = P\{ |N(0,1)| > |X_i| \}$. After sorting the $p$-values in
ascending order $p_{(1)} < p_{(2)} < \ldots < p_{(n)}$, we define the
$HC$ at each pixel with $p$-value $p_i$, by:
$$
HC_{n,i}  =    \sqrt{n} \biggl|  \frac{i/n  - p_{(i)}}{\sqrt{p_{(i)}
(1-p_{(i)})}}  \biggr|,
$$

We compute the values of the HC statistic of the 3--year WCM in real and in wavelet space.
The obtained values of the HC statistic 
are presented in Figure~\ref{fig:HC_1}.  These values correspond to the maximum of
the HC values found at the individual pixels. As in previous figures, 
circles denote the results obtained from 
the 3--year WCM, asterisks those from the 1--year WCM
and the bands represent the acceptance intervals. 
As one can see in the Figure, the data in wavelet space are not
compatible with Gaussian predictions at scales $R_8$ and $R_9$ at the $99\%$ c.l. 
This is in agreement with the result obtained by CJT for the 1--year WMAP data although there the
HC values at scale $R_8$ were just below the $99\%$ c.l. 
The upper tail probabilities for the 1--year and 3--year maximum HC values at scale $R_9$, 
are 0.56\% and 0.36\% respectively.
The map of HC values at scale $R_9$ is presented in 
Figure~\ref{fig:HC_2}. It is clear that the pixels responsible for the detected deviation
from Gaussianity are located at the position of \emph{the Spot}. Convolution with 
the wavelet causes the observed ring structure in the HC map. Figure~\ref{fig:HC_3} 
shows a blowout image of \emph{the Spot} as it appears at scale $R_9$ 
in the wavelet map and in the HC map.   

\section{Significance}

In the previous section, the upper tail probabilities of each
estimator at scale $R_9$ were given. All the considered estimators showed the lowest upper
tail probability at scale $R_9$.
However these are not rigorous measures of the
significance of \emph{the Spot}, since the number of performed tests is not
taken into account.
In this section we will recalculate the $p$-value of the deviation in the kurtosis found 
by V04 and discuss the issue of \emph{a posteriori} significances.

When an anomaly is detected in a data set following a blind approach,
usually several additional tests are performed afterwards to further
characterize the anomaly. In most of these cases, the only reason these tests
have been performed is the previous finding of the initial anomaly. 
If another anomaly would have been detected, other followup
tests would have been performed.
Hence these followup tests have not been performed blindly and should
not be taken into account to calculate the significance of the initial
detection.

This issue was already discussed in C06 and McEwen et
al. (2005). Both papers recalculated the significance of the
excess of kurtosis in the 1--year WCM found by V04. 
The excess of kurtosis was found performing a blind test, since no model was used and 
no previous findings conditioned the choice of the scales.
Since 15 wavelet
scales and two estimators (skewness and kurtosis) were considered,
a total sum of 30 tests were performed. 
Three of these tests detected a strong deviation from Gaussianity. Scales $R_7$, $R_8$ and $R_9$ 
presented upper tail probabilities 0.67\%, 0.40\% and 0.38\% in the 1--year data.
This fact was taken into account in C06, but it was not by McEwen et al. (2005). 
The latter searched through the simulations in 
order to find how many of them showed a higher or equal deviation than 
the maximum deviation of the data, ignoring that the data showed a 
high deviation at two adjacent scales. The $p$-value found 
in this way was 4.97\% whereas C06 obtained 1.91\% taking into account 
that the data deviate at three consecutive scales. 
It is also interesting to note that, when both Galactic hemispheres were considered
independently, C06 found a $p$-value of 0.69\%, although this could be considered as a followup test.

Some readers could find that the three-consecutive-scales criterion is an \emph{a posteriori} choice since
we look first at the data and given that they deviate at three consecutive scales, we then calculate from 
the simulations how probable this is.
Therefore we should consider a new test which eliminates this \emph{a posteriori} choice.
We fix \emph{a priori} a significance level which is the 1\% acceptance interval given in all figures, 
and count for each estimator (skewness and kurtosis) how many scales lie outside, no matter if they are consecutive or not. 
Then we search through the simulations how many show at least that many scales outside the 1\% acceptance 
interval as the data.

Applying this test to the 3--year WCM, we find that scales $R_8$ and $R_9$ lie outside the 1\% acceptance 
interval and scale $R_7$ lies on the border for the kurtosis estimator as can be seen in 
Figure~\ref{fig:kurtosis}. Searching through the simulations
how many deviate in three scales either in the skewness or in the kurtosis estimator, we find a 
$p$-value of 1.85\%, which is still below the $p$-value obtained for the 1--year data with 
the three-consecutive-scales criterion.

As already discussed we should not include the followup tests in a rigorous significance analysis.
However it is difficult to assess if some of these tests would have been performed or not without
the first finding of V04. In fact, the area and maxima analyses are very intuitive and simple.
If V04 had performed their blind analysis on those estimators instead of using skewness and kurtosis,
then the significance would be different. 
We should distinguish between those tests which are clearly followup tests, because the only reason they have 
been performed is the initial
detection, and other tests which just have been performed after the initial detection, but could have been
performed before.

Hence we apply our new robustness test to kurtosis, Max, Area at thresholds 3.0 and 4.0 and Higher Criticism separately.
Note that whereas the first two estimators are two-sided, the Area and Higher Criticism are one sided estimators.
The $p$-values obtained in this way are listed in Table~\ref{table:pvalues}.
The kurtosis and Area at threshold 4.0 show $p$-values around 1\%, Higher Criticism and Area at threshold 3.0 around 3\%.
On the contrary the Max estimator does not show a significant deviation from Gaussianity according to this robustness test.

\begin{deluxetable}{ccc}
\tablewidth{0pt}
\tablecaption{$p$-values for different estimators.\label{table:pvalues}}
\tablecolumns{2}
\tablehead{ Estimators & $p$-value }
\startdata
	  kurtosis & 0.86\% \\ 
	  skewness + kurtosis & 1.85\% \\ 
	  Max  & 11.64\% \\ 
	  Area 3.0  & 3.27\% \\ 
	  Area 4.0  & 1.09\% \\ 
	  Higher Criticism & 3.48\% \\ 
\enddata
\end{deluxetable}

The most conservative and reliable value is the 1.85\% figure since it is not suspicious of being obtained through 
\emph{a posteriori} analyses.
Nevertheless it is still noticeable that the followup tests performed in C05, C06, CJT and in the present 
paper, confirm the initial finding with a very similar significance. 
Even if strictly speaking these should not be taken into account for
establishing the significance of \emph{the Spot}, they confirm the robustness of the detection.
\begin{figure}
  \begin{center}
    \includegraphics[width=84mm,height=84mm]{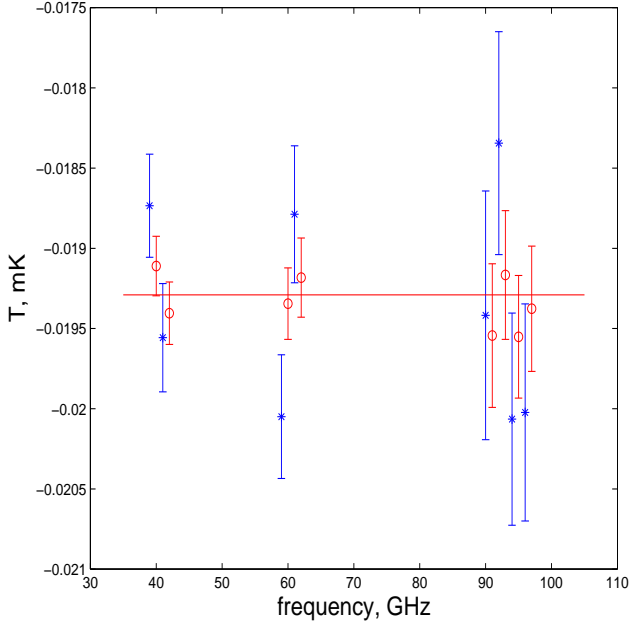}
  \end{center}
  \caption{Frequency dependence of the temperature at the center of \emph{the Spot} at scale $R_9$.
    Again the asterisks represent the 1--year data and the circles the 3--year data. The horizontal
    line shows the value of the 3--year WCM.
    The data at the same frequency have been slightly offset in abscissa for readability.}
  \label{fig:T_freq}
\end{figure}
\begin{figure}
  \begin{center}
    \includegraphics[width=84mm]{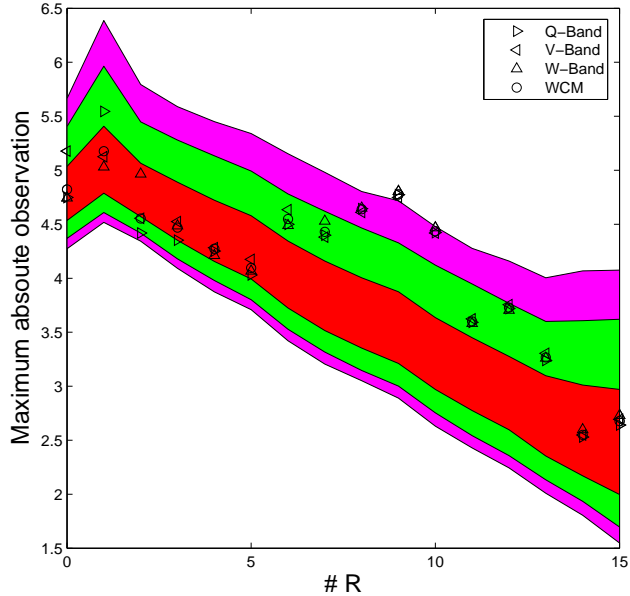}
  \end{center}
  \caption{Maximum absolute observation for the Q, V and W bands, compared to the 3--year WCM values.}
  \label{fig:Max_freq}
\end{figure}
\begin{figure}
  \begin{center}
    \includegraphics[width=84mm]{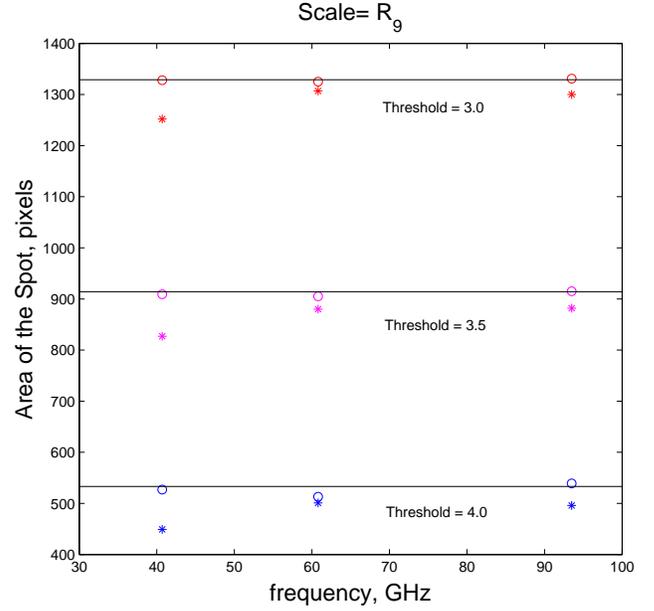}
  \end{center}
  \caption{Frequency dependence of the Area of \emph{the Spot} at scale $R_9$ and several 
    thresholds. Asterisks represent the 1--year data and the circles the 3--year data. The 3--year WCM values 
    are represented by horizontal lines.}
  \label{fig:Area_freq}
\end{figure}
\begin{figure}
  \begin{center}
    \includegraphics[width=84mm]{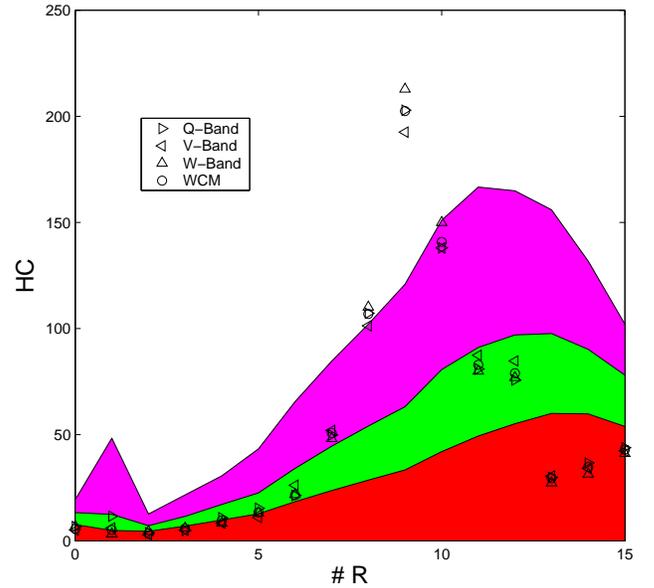}
  \end{center}
  \caption{Higher Criticism values for the Q, V and W bands, compared to the 3--year WCM values.}
  \label{fig:HC_freq}
\end{figure}

\section{Frequency dependence}

\begin{figure}
  \begin{center}
    \includegraphics[width=84mm,height=84mm]{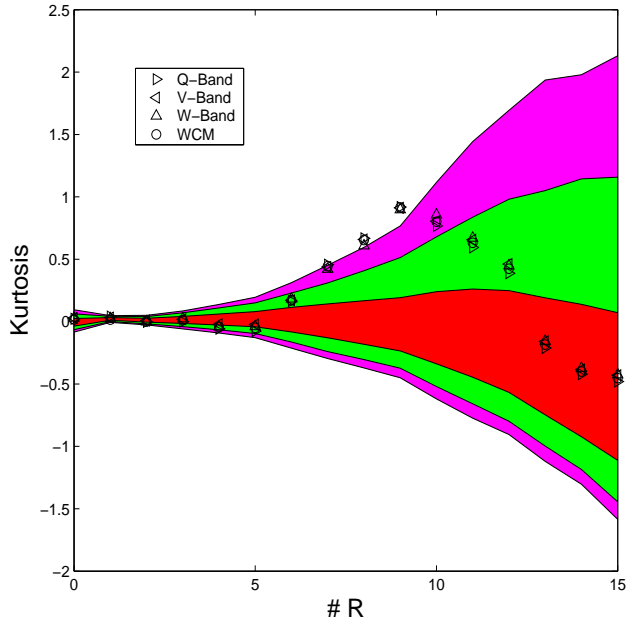}
  \end{center}
  \caption{Kurtosis values for the Q, V and W bands, compared to the 3--year WCM values.}
  \label{fig:kurt_QVW} 
\end{figure}

In this section we will analyse the frequency dependence of the previously analysed estimators.
A flat frequency dependence is characteristic of CMB, whereas other emissions such as Galactic
foregrounds show a strong frequency dependence.
Figure~\ref{fig:kurt_QVW} shows that the kurtosis has almost identical values at the three foreground 
cleaned channels, namely Q,V and W. Same behaviour was observed in the 1--year data (see Figure 7 in C06). 
Strong frequency dependent foreground emissions are unlikely to produce the detected
excess of kurtosis.

The frequency dependence of the temperature at the center of
\emph{the Spot}, i.e. at the pixel where the temperature of \emph{the Spot} is minimum in the 
WCM map, is presented in Figure~\ref{fig:T_freq}.
The error bars of the 1--year data have been estimated performing 1000 noise simulations as 
explained in section 5.1 of C06. As the noise variance is $\approx 3$ times lower in the
3--year data, we estimate the new error bars simply by dividing the old ones by $\sqrt{3}$.
No frequency dependence is found for the new data set
in agreement with the results for the 1--year data. 
$Max$, Area and HC values at different frequencies (see Figure~\ref{fig:Max_freq}, 
Figure~\ref{fig:Area_freq} and Figure~\ref{fig:HC_freq}) show a very low relative variation 
compared to the 3--year WCM.

All these results confirm the analysis performed in section 5 of C06 where the data 
were found to fit a flat 
CMB spectrum. The present analysis confirms the disagreement between the conclusions of C06 and those of the work of Liu \& Zhang (2005) where Galactic foregrounds were
considered to be the most likely source for non-Gaussian features found with spherical wavelets.

\section{Discussion}

Spergel et al. (2006) enumerate several reasons to be cautious about the different anomalies found in the WMAP data:
Galactic foregrounds or noise could be generating the non-Gaussianity, and in addition most of
the claimed detections are based on \emph{a posteriori} statistics. Also spatial variations of the noise variance and 
$1/f$ noise could affect some of the perfomed analyses. They suggest several tests to be
done using difference maps (year 1 - year 2, year 2 - year 3, etc.) and multi-frequency data.

We have tried to address all those points for \emph{the Spot}.
The \emph{a posteriori} analysis is one of the most important issues raised by Spergel et al. (2006),
since it is very difficult to get completely rid of it. Most analyses perform many tests and it is not
easy to assess how many of them are followup tests and which is the probability of finding an anomaly
by chance.
As discussed in section 4 a very careful analysis shows that \emph{the Spot} remains statistically 
significant at least at the 98\% confidence level, without using any \emph{a posteriori} statistics.

In addition C06 proved that \emph{the Spot} 
remained highly significant no matter which foreground reduction technique
was used. These results are confirmed in the present paper. The new foreground reduction used in
the 3--year data enhances slightly the significance of our detection.
Moreover the multi-frequency analysis of the previous section shows an even flatter frequency 
dependence of \emph{the Spot}. 

As already discussed in previous sections the noise does not affect significantly our wavelet analysis.
In fact the coadded 3--year results are very similar to those obtained with the 1--year data of the new data release. 
No significant cold spot is observed based on the analysis of the three difference maps (year 1 - year 2, year 2 - year 3, and year 1 - year 3).
Moreover Figure~\ref{fig:T_freq} shows that even the particularly $1/f$ contaminated W4 Difference Assembly shows almost the same result
as all the other Difference Assemblies.

\section{Conclusions}

In this paper we repeat the analyses that detected the non-Gaussian cold spot called \emph{the Spot} at ($b = -57^\circ, l = 209^\circ$) in wavelet space 
in the 1--year of WMAP data, using the recently released 3--year WMAP data. 
The previous 
works V04, C05, CJT and C06 found \emph{the Spot} to deviate significantly from the Gaussian 
behaviour. \emph{The Spot} was detected using several estimators,
namely kurtosis, Area, $Max$ and HC. This work confirms the detection applying all these 
estimators to the recently published 3--year WMAP data. 
At scale $R_9$, the upper tail probabilities
of all these estimators when applied to the 3 year WMAP data are smaller
than the corresponding ones for the first year WMAP data. This is mostly due to the improved foreground reduction of the data.
We calculate the probability of finding such a deviation from Gaussianity considering only skewness and kurtosis
since these were initially used by V04 following a blind approach. Therefore excluding followup tests
which could be considered as \emph{a posteriori} analyses we obtain a $p$-value of $1.85\%$.
Moreover, \emph{the Spot} appears to be almost 
frequency independent. This result  reinforces the previous foreground analyses performed by C06. 
It is very unlikely that foregrounds are responsible for the non-Gaussian behaviour
of \emph{the Spot}.
Comparing the WMAP single year sky maps, we conclude that the noise has a very low contribution to our
wavelet analysis as already claimed in V04, C05.
Future works will be aimed at finding the origin of \emph{the Spot}. As discussed in the 
introduction several possibilities have been considered, based on Rees-Sciama effects (Rees \& Sciama 1968, 
Mart{\'\i}nez--Gonz{\'a}lez \& Sanz 1990, Mart{\'\i}nez--Gonz{\'a}lez et al. 1990) and inhomogenous or 
anisotropic universes.
We are presently working on studying another: topological defects (Turok \& Spergel 1990, Durrer et al. 1999) as textures could produce cold spots.
New and more detailed analyses are required in order to answer that question.

\begin{acknowledgements}
The authors kindly thank R.B. Barreiro, L.M. Cruz-Orive and J.L. Sanz for very useful comments and R. Marco for
computational support.
MC thanks Spanish Ministerio de Educacion Cultura y Deporte (MECD) for a predoctoral FPU fellowship.
PV thanks a I3P contract from the Spanish National Research Council (CSIC).
MC, EMG and PV acknowledge financial support from the Spanish MCYT project ESP2004-07067-C03-01 and 
the use of the Legacy Archive for Microwave Background Data 
Analysis (LAMBDA). Support for LAMBDA is provided by the NASA Office of Space 
Science.
This work has used the software package HEALPix (Hierarchical, Equal
Area and iso-latitude pixelization of the sphere,
http://www.eso.org/science/healpix), developed by K.M. G{\'o}rski,
E. F. Hivon, B. D. Wandelt, J. Banday, F. K. Hansen and
M. Barthelmann; the visualisation program Univiewer, developed by S.M. Mingaliev, M. Ashdown and 
V. Stolyarov; and the CAMB and CMBFAST software, developed by  A. Lewis and A. Challinor and 
by U. Seljak and M. Zaldarriaga respectively.
\end{acknowledgements}


\begin{thebibliography}{}
%
\bibitem{adl06} Adler R. J., Bjorken J. D., Overduin J. M., 2006, gr-qc/0602102.
%
\bibitem{ben03a} Bennett C.L., et al., 2003, ApJS, 148, 1.
%
\bibitem{ben03b} Bennett C.L., et al., 2003, ApJS, 148, 97.
%
\bibitem{bielewicz:2005} Bielewicz P., Eriksen H. K., Banday A. J., G\'{o}rski K. M., Lilje P. B., 2005, ApJ, 635, 750B 
%
\bibitem{cab05} Cabella P., Liguori M., Hansen F.K., Marinucci D.,Matarrese S., Moscardini L., Vittorio N., 2005, MNRAS, 358, 684.
%
\bibitem{cay05} Cay\'on L., Jin J., Treaster A., 2005, MNRAS, 362, 826.(CJT)
%
\bibitem{cay06} Cay\'on L., Banday A. J., Jaffe T., Eriksen H. K., Hansen F.K., Gorski K. M., Jin J., 2005, submitted to MNRAS, (astro-ph/0602023).
%
\bibitem{chiang:2003} Chiang L. -Y., Naselsky P. D., Verkhodanov O. V., 2003, ApJ, 590, 65.
%
\bibitem{chi06} Chiang L-Y, Naselsky P.D., Coles P., 2006, (astro-ph/0603662).
%
\bibitem{chyzy05} Chyzy K.T., Novosyadlyj B., Ostrowski M., 2005, (astro-ph/0512020).
%
\bibitem{coles:2004} Coles P., Dineen P., Earl J., Wright D., 2004, MNRAS, 350, 989.
%
\bibitem{copi:2004} Copi C. J., Huterer D., Starkman G. D., 2004, Phys. Rev. D., 70, 043515.
%
\bibitem{copi:2005} Copi C. J., Huterer D., Schwarz D. J., Starkman G. D., 2005, submitted to MNRAS (astro-ph/0508047).
%
\bibitem{cru05} Cruz M., Mart\'{\i}nez--Gonz\'alez E., Vielva P., Cay\'on L., 2005, MNRAS, 356, 29. (C05).
%
\bibitem{cru06} Cruz M., Tucci M., Mart\'{\i}nez--Gonz\'alez E., Vielva P., 2006, accepted in MNRAS, (astro-ph/0601427). (C06).
%
\bibitem{dineen:2005} Dineen P., Coles P., 2005, submitted to MNRAS (astro-ph/0511802).
%
\bibitem{Don:2004} Donoho D., Jin J., 2004, {\it Ann. Statist.}, 32, 962 
%
\bibitem{dur99} Durrer R., 1999, New Astron. Rev., 43, 111.
%
\bibitem{eriksen:2005} Eriksen H. K., Banday A. J., G\'{o}rski K. M., Lilje P. B., 2005, ApJ, 622, 58.
%
\bibitem{eriksen:2004a} Eriksen H. K., Hansen F. K., Banday A. J., G\'{o}rski K. M., Lilje P. B., 2004, ApJ, 605, 14.
%
\bibitem{eriksen:2004b} Eriksen, H. K., Novikov, D. I., Lilje, P. B, Banday, A. J., G\'{o}rski K. M., 2004, ApJ, 612, 64.
%
\bibitem{Gor05} G\'{o}rski K.M., Hivon E. F., Wandelt B. D., Banday J., Hansen F. K., Barthelmann M., 2005, ApJ 622, 759.
%
\bibitem{hansen:2004} Hansen F.K., Cabella P., Marinucci D., Vittorio N., 2004, ApJL, 607, L67.
%
\bibitem{hinshaw:2006} Hinshaw et al., 2006, submitted to ApJ, (astro-ph/0603451)
%
\bibitem{inoue06} Inoue K. T., Silk J., 2006, (astro-ph/0602478)
%
\bibitem{jaffe:2005a} Jaffe T. R., Banday A. J., Eriksen H. K., G\'{o}rski K. M., Hansen F. K., 2005, ApJ, 629, 1.
%
\bibitem{jaffe:2005b} Jaffe T. R., Hervik S., Banday A. J., G\'{o}rski K. M., 2005, submitted to ApJ (astro-ph/0512433).
%
\bibitem{komatsu:2003} Komatsu E. et al. , 2003, ApJs, 148, 119.
%
\bibitem{lm:2005a} Land K., Magueijo J., 2005a, MNRAS, 357, 994.
%
\bibitem{lm:2005b} Land K., Magueijo J., 2005b, MNRAS, 362, 16. 
%
\bibitem{lm:2005c} Land K., Magueijo J., 2005d, MNRAS, 362, 838. 
%
\bibitem{larson:2004} Larson D. L., Wandelt B. D., 2004, ApJ, 613, 85.
%
\bibitem{larson:2005} Larson D. L., Wandelt B. D., 2005, submitted to Phys. Rev. D. (astro-ph/0505046).
%
\bibitem[{Liu \& Zhang}{2005}]{liu05} Liu X., Zhang S.N., 2005, ApJ, 633, 542.
%
\bibitem{Mart1990a} Mart{\'\i}nez--Gonz{\'a}lez E. \& Sanz J. L., Silk, J., 1990, ApJL, 335, 5.
%
\bibitem{Mart1990b} Mart{\'\i}nez--Gonz{\'a}lez E. \& Sanz J. L., 1990, MNRAS, 247, 473.
%
\bibitem{SMHW-SHW:PLANCK} Mart{\'\i}nez--Gonz{\'a}lez E., Gallegos J. E., Arg{\"u}eso F., Cay{\'o}n L. \& Sanz J. L., 2002, MNRAS, 336, 22.
%
\bibitem{mcewen:2005a} McEwen J. D., Hobson M. P., Lasenby A. N., Mortlock D. J., 2005, MNRAS, 359, 1583.	
%
\bibitem{mcewen:2005b} McEwen J. D., Hobson M. P., Lasenby A. N., Mortlock D. J., 2005, submitted to MNRAS (astro-ph/0510349).	
%
\bibitem{mw:2004} Mukherjee P., Wang Y., 2004, ApJ, 613, 51.
%
\bibitem{oliveira:2004} de Oliveira-Costa A., Tegmark M., Zaldarriaga M., Hamilton A., 2004, Phys. Rev. D., 69, 63516.
%
\bibitem{Park2004} Park C. G. 2004, MNRAS 349, 313-320.
%
\bibitem{RS:1968} Rees M. J. \& Sciama D. W., 1968, Nature, 517, 611.
%
\bibitem{schwarz:2004} Schwarz D. J., Starkman G. D., Huterer D., and Copi C. J.,  2004, Phys. Rev. Lett., 93, 221301.
%
\bibitem{slosar:2004} Slosar A., Seljak U., 2004, Phys. Rev. D., 70, 8. 
%
\bibitem{spergel:2006} Spergel et al. 2006, submitted to ApJ, (astro-ph/0603449)
%
\bibitem{toj05} Tojeiro R., Castro P.G., Heavens A.F., Gupta S., 2005, submitted to MNRAS (astro-ph/0507096).
%
\bibitem{tom05} Tomita K., 2005, Phys. Rev. D, 72, 10.
%
\bibitem{turok90} Turok N. \& Spergel D. N., 1990, Phys. Rev. Letters, 64, 2736
%
\bibitem{vie04} Vielva P., Mart\'{\i}nez--Gonz\'alez E., Barreiro R. B., Sanz J.L., Cay\'on L., 2004, ApJ, 609, 22. (V04).
%
\bibitem{wiaux:2006} Wiaux Y., Vielva P., Mart\'{\i}nez--Gonz\'alez E., Vandergheynst P., 2006, submitted to Phys. Rev. Letters.
\end{thebibliography}
\end{document}